\documentclass[sigconf]{acmart}

\usepackage{pbox}
\usepackage{xcolor}
\usepackage[utf8]{inputenc}
\usepackage{tabu}
\usepackage{caption}
\usepackage{subcaption}
\usepackage[para]{footmisc}
\usepackage{adjustbox}
\usepackage{balance}
\usepackage[show]{chato-notes}
\usepackage{multirow}
\usepackage{tablefootnote}
\usepackage[noadjust]{marginnote}

\AtBeginDocument{%
  }

\newcolumntype{H}{>{\setbox0=\hbox\bgroup}c<{\egroup}@{}}
\newcolumntype{Z}{>{\setbox0=\hbox\bgroup}c<{\egroup}@{\hspace*{-\tabcolsep}}}
\newcommand{\pageenlarge}[1]{\marginnote{}\enlargethispage{#1\baselineskip}}

\newcommand{\zl}[1]{\textcolor{black}{#1}}

\newcommand{\zz}[1]{\textcolor{black}{#1}}

\newcommand{\crp}[1]{\textcolor{black}{#1}}

\copyrightyear{2025}
\acmYear{2025}
\setcopyright{cc}
\setcctype{by}
\acmConference[SIGIR '25]{Proceedings of the 48th International ACM SIGIR Conference on Research and Development in Information Retrieval}{July 13--18, 2025}{Padua, Italy}
\acmBooktitle{Proceedings of the 48th International ACM SIGIR Conference on Research and Development in Information Retrieval (SIGIR '25), July 13--18, 2025, Padua, Italy}\acmDOI{10.1145/3726302.3730179}
\acmISBN{979-8-4007-1592-1/2025/07}

\begin{document}

\title{Document Similarity Enhanced IPS Estimation for Unbiased Learning to Rank}

\author{Zeyan Liang}
\affiliation{%
  \institution{University of Glasgow}
  \city{Glasgow}
  \country{Scotland, UK}
  \postcode{G12 8QQ}
}
\email{z.liang.1@research.gla.ac.uk}

\author{Graham McDonald}
\affiliation{%
  \institution{University of Glasgow}
  \city{Glasgow}
  \country{Scotland, UK}
  \postcode{G12 8QQ}
}
\email{graham.mcdonald@glasgow.ac.uk}

\author{Iadh Ounis}
\affiliation{%
  \institution{University of Glasgow}
  \city{Glasgow}
  \country{Scotland, UK}
  \postcode{G12 8QQ}
}
\email{iadh.ounis@glasgow.ac.uk}
\begin{abstract}
Learning to Rank (LTR) models learn from historical user interactions, such as user clicks. However, there is an inherent bias in the clicks of users due to position bias, i.e., users are more likely to click highly-ranked documents than low-ranked documents. To address this bias when training LTR models, many approaches from the literature re-weight the users' click data using Inverse Propensity Scoring (IPS). IPS re-weights the user's clicks proportionately to the position in the historical ranking that a document was placed when it was clicked since low-ranked documents are less likely to be seen by a user. In this paper, we argue that low-ranked documents that are similar to highly-ranked relevant documents are also likely to be relevant. Moreover, accounting for the similarity of low-ranked documents to highly ranked relevant documents when calculating IPS can more effectively mitigate the effects of position bias. Therefore, we propose an extension to IPS, called IPSsim, that takes into consideration the similarity of documents when estimating IPS. We evaluate our IPSsim estimator using two large publicly available LTR datasets under a number of simulated user click settings, and with different numbers of training clicks. Our experiments show that our IPSsim estimator is more effective than the existing IPS estimators for learning an unbiased LTR model, particularly in top-n settings when n $>=$ 30. For example, when n = 50, our IPSsim estimator achieves a statistically significant $\sim$3\% improvement ($p < 0.05$) in terms of NDCG compared to the Doubly Robust estimator from the literature.
\end{abstract}



\maketitle

\section{Introduction}
Learning to Rank (LTR)~\cite{liu2009learning} models are an effective strategy for generating relevant search results in response to a user's query. Such LTR models are trained using the interaction data, e.g., \zz{the} clicks, of historical users as a signal of relevance~\cite{joachims2017unbiased}. However, there is an inherent bias in the clicks of historical users due to the position bias~\cite{joachims2017accurately,joachims2002optimizing,agarwal2014reliable}, i.e., the fact that users are more likely to click documents that are ranked higher in the search results compared to lower-ranked documents. Moreover, highly-ranked documents tend to receive more clicks than low-ranked documents, regardless of the relevance of the documents~\cite{joachims2017unbiased}.

Training LTR models on data that is skewed by position bias can lead to these biases being propagated by the learned model, and the biases can be exaggerated in the search results that are generated by the newly trained model~\cite{craswell2008experimental}. To address this inherent bias and mitigate its effects, many approaches from the literature, e.g.,~\cite{joachims2017unbiased,agarwal2019addressing,ai2021unbiased,oosterhuis2023doubly}, re-weight users' click data using Inverse Propensity Scoring (IPS)~\cite{joachims2017unbiased} when training a LTR model. This is typically referred to as Unbiased Learning to Rank (ULTR)~\cite{agarwal2019general}.

\looseness -1 IPS corrects for position bias by re-weighting clicks based on the inverse of their propensity (i.e., the probability of being observed at a given rank)~\cite{joachims2017unbiased}. However, relying solely on position may \zz{fail to capture} the true relevance of the documents \zz{that are} ranked low and \zz{are} thus rarely clicked. Prior studies have shown that documents with high document similarity tend to share the same relevance labels \cite{macavaney2022adaptive}. Moreover, Saito~\cite{saito2020unbiased} showed that the similarity of items can help to mitigate position bias in recommender systems. Building on these insights, we hypothesise that such low-ranked relevant documents are likely to be topically and/or semantically similar to the highly-ranked relevant documents. Moreover, the similarity between low-ranked documents and highly-ranked documents can potentially be leveraged when calculating IPS to mitigate position bias. 

In this work, we propose an extension to IPS, namely IPSsim, that adjusts the IPS calculation to consider the similarity between low-ranked documents and highly-ranked documents. We evaluate our IPSsim estimator using two large publicly available LTR datasets, ISTELLA-S~\cite{dato2016fast} and MSLTR-WEB30K~\cite{qin2013introducing}, which are standard ULTR benchmarks. Moreover, we also evaluate the effectiveness of our IPSsim estimator when it is instantiated with three different similarity measures. Our experiments show that our IPSsim estimator is more effective at mitigating position bias in historical rankings than existing IPS estimators, particularly in top-n settings where n >= $30$. For example, when n = 50 and the number of training clicks is \( 10^6 \), our IPSsim estimator achieves a statistically significant 2.88\% improvement ($p < 0.05$) in terms of NDCG compared to the Doubly Robust~\cite{oosterhuis2023doubly} (DR) estimator on ISTELLA-S dataset. 

\section{Related Work}
\zz{Users' interactions with search systems can be influenced by various biases, such as trust bias~\cite{joachims2007evaluating}, item selection bias~\cite{marlin2009collaborative}, popularity bias~\cite{canamares2018should}, and others~\cite{prosperi2020causal,yue2010beyond,moscovici1972social}.} In this work, we are interested in mitigating the effects of position bias~\cite{joachims2002optimizing}, where documents that are presented at the top of a ranking receive disproportionately more clicks regardless of the true relevance of the documents. Most previous ULTR work uses (semi-)synthetic datasets to control for the interaction effects of different biases, e.g.,~\cite{gupta2023safe,joachims2017unbiased,vardasbi2020inverse,oosterhuis2020policy}. \zz{We follow the experimental setup of Oosterhuis~\cite{oosterhuis2023doubly} and use the same semi-synthetic datasets in our experiments.}

Recent ULTR models typically deploy mitigation strategies based on causal inference to \crp{debias} the click data before training the LTR model. For example, IPS~\cite{wang2016learning} is one mitigation strategy that has been shown to be particularly effective. Wang et al.~\cite{wang2016learning} first proposed IPS as a method to mitigate query selection bias for LTR in the context of personal search. Joachims et al.~\cite{joachims2017unbiased} then generalised the use of IPS in LTR and demonstrated, both theoretically and empirically, the effectiveness of IPS for \crp{debiasing} user click data for ULTR. Vardasbi et al.~\cite{vardasbi2020inverse} proposed an affine correction for IPS to further correct for trust bias in user clicks. In this work, we extend Vardasbi’s IPS definition~\cite{vardasbi2020inverse,oosterhuis2023doubly} to also incorporate the similarity of low-ranked documents to highly-ranked relevant documents when calculating IPS. In our experiments, we also deploy the IPS variant from Vardasbi et al.~\cite{vardasbi2020inverse} \zl{as one of our baseline approaches for comparison with our proposed IPSsim approach.}

IPS methods, e.g.,~\cite{joachims2017unbiased, ai2018unbiased, oosterhuis2023doubly}, can reduce the bias in estimations of click probabilities. However, when there is insufficient available click data, high variance in IPS methods can lead to ULTR performing sub-optimally~\cite{wang2016learning,gupta2023safe,oosterhuis2023doubly,wang2021non}. To address this, \zl{Wang et al.~\cite{wang2021non} introduced the Ratio-Propensity Scoring (RPS) estimator. Variance in IPS estimation} has also been addressed by the Direct Method (DM) and Doubly Robust (DR) estimators proposed by Oosterhuis~\cite{oosterhuis2023doubly}. DR is a more robust and more effective version of DM. Therefore, in this work, we also compare our IPSsim estimator against the RPS and DR estimators.

Saito et al.~\cite{saito2022off} proposed the \crp{Marginalised} IPS (MIPS) estimator, which used categorical side information (e.g., movie genres) as a measure of similarity to adjust an item's IPS to better align with the click-through rates of similar items. In this work, we compare our proposed IPSsim approach with MIPS. \zz{However,} since side information is \zz{typically not} available in an ULTR context, we compare against two versions of MIPS that are adapted for ULTR, namely \mbox{MIPS\textsubscript{LTR}} and Doubly Robust \crp{Marginalised} IPS (DRM). We provide details of \mbox{MIPS\textsubscript{LTR}} and DRM in Section~\ref{sec:setup}.

\section{Proposed Approach}\label{sec:approach}
\pageenlarge 1
Given a set of historical rankings, $h \in H$, for a query, $q$, and the corresponding biased user clicks, $c \in C$, our proposed IPSsim estimator mitigates the position bias in the historical click data. To do this, we hypothesise that low ranked documents that are similar to highly ranked relevant documents are likely to also be relevant. Therefore, we leverage the similarity of a document, $i \in I$, where $I$ is the set of all documents in $H$, to the documents, denoted as $T$, that are highly ranked in $H$. With this in mind, we first define our 
\begin{equation}
    \text{Avgsim}(i,T) = \frac{1}{|T|} \sum_{i_m \in T, i_m \neq i} \text{sim}(i, i_m)
\label{eq:average_sim}
\end{equation}
document similarity function, Avgsim($i,T$), as above. where \( \text{sim}(i, i_m) \) calculates the similarity between documents \( i \) and \( i_m \), and $|T|$ is the number of highly ranked documents. Avgsim($i,T$) calculates for each document, $i \in I$, how similar $i$ is to $T$, in other words the average similarity between the document $i$ and the top-ranked documents in $H$. To integrate the document similarities from Eq.~\ref{eq:average_sim} into the IPS calculation, we define our proposed IPSsim estimator as follows:
\begin{equation}
     \resizebox{1\linewidth}{!}{$
    {\text{IPSsim}}(H,C) = \\ \frac{1}{|H|} \sum_{h=1}^{|H|} \sum_{i \in I} \hat{\omega}_i \cdot \left(\frac{(1-\alpha)}{\hat{\rho}i} + \alpha \cdot \text{Avgsim}(i,T)\right) (c_h(i) - \hat{\epsilon}_{k_h}(i))
     $}
\label{eq:ipssim}
\end{equation}
\looseness -1 where $\hat{\omega_i}$ is the probability of the document $i$ appearing at position $k$ under the target policy and $\hat{\rho}i$ is the propensity score of document $i$ at position $k$ in historical ranking $h$, from the original IPS formulation $\frac{\hat{\omega}_i}{\hat{\rho}i}$~\cite{vardasbi2020inverse, oosterhuis2023doubly}. The variable $\alpha$ in the range $[0..1]$ provides a weighting factor to control for the relative contribution from the original IPS component and the Avgsim$(i, T)$ component of IPSsim. Lastly, $c_h(i)$ is a binary click/no-click observation for document $i$ in a specific ranking $h$ and $\hat{\epsilon}_{k_h}(i)$ is a trust bias correction~\cite{vardasbi2020inverse, oosterhuis2023doubly} term for a document at position $k$.    

\section{Experimental Setup}\label{sec:setup}
\pageenlarge 1
We conduct experiments to investigate the following three research questions: 
\textbf{RQ1}: Does incorporating document similarity scores into IPS estimations improve the effectiveness of IPS? 
\textbf{RQ2}: Does the top-n setting affect the performance of our IPSsim estimator? 
\textbf{RQ3}: How does the number of training clicks affect the overall performance of the IPSsim estimator?

\looseness -1\noindent \textbf{Datasets:} To comprehensively evaluate our IPSsim estimator, following~\cite{oosterhuis2023doubly}, we simulate user clicks for two commercial search datasets, namely (1) ISTELLA-S~\cite{dato2016fast} with 33,018 queries, 3,408,630 documents and 220 feature values, and (2) the Microsoft learning to rank dataset MSLTR-WEB30K with 30,000 queries, 31,531,961 documents and 136 feature values~\cite{qin2013introducing}. Both datasets are LETOR formatted with expert-annotated relevance labels, $rel(i) \in \{0, 1, 2, 3, 4\}$. We use the provided training, validation, and test splits. 

\noindent \textbf{Baselines:} We compare our IPSsim estimator against the following methods. Firstly, the Naive approach~\cite{ai2021unbiased,oosterhuis2023doubly} that trains \zz{a} LTR model using simulated clicks without any bias correction. We also compare against RPS~\cite{wang2021non} and the IPS estimator from Vardasbi et al.~\cite{vardasbi2020inverse}, as well as the DR estimator proposed by Oosterhuis~\cite{oosterhuis2023doubly}. Additionally, we adapt MIPS~\cite{saito2022off} to our ULTR setting. 

When computing IPS for a document $i$, MIPS~\cite{saito2022off} leverages item similarity (e.g., genre of movies). In our ULTR setting, we lack such side information, so we calculate cosine similarity~\cite{singhal2001modern} by features instead. Specifically, for each historical ranking $h$, our variant MIPS\textsubscript{LTR} defines the importance weight $\hat{\omega}_i$ as: $\hat{\omega}_i = \frac{1}{s} \sum_{j=1}^{s} M\bigl(k \mid i_j\bigr)\,\mathrm{sim}\bigl(i, i_j\bigr)$, where $M(k \mid i_j)$ is the probability that document $i_j$ is ranked at position $k$ in $h$, and $\mathrm{sim}(i, i_j)$ is the Cosine similarity between documents $i$ and $i_j$. Here, $s$ is the total number of documents in $h$ excluding $i$. The final MIPS weight is adjusted by dividing $\hat{\omega_i}$ by the propensity score $\hat{\rho}_i$, and the MIPS estimate is computed as:

\begin{equation}
    MIPS_{LTR} = \frac{1}{|H|} \sum_{h=1}^{|H|} \sum_{i \in I} \left( \frac{\hat{{\omega}_i}}{\hat{\rho}_i} \right) \left( c_h(i) - \hat{\epsilon}_{k_h(i)} \right)
\label{eq:mips}
\end{equation}

Our DRM variant combines MIPS\textsubscript{LTR} with the DR approach proposed by Oosterhuis et al.~\cite{oosterhuis2023doubly}, by replacing DR's IPS component with MIPS\textsubscript{LTR}. Note that for both MIPS\textsubscript{LTR} and DRM, differently from MIPS~\cite{saito2022off}, all documents have some similarity to $i$ and are, therefore, included in the set of $s$ documents that are similar to $i$.

\noindent\textbf{IPSsim Parameters:} We evaluate three similarity measures for \(\text{Avgsim}(i_k)\) in Eq.~\ref{eq:average_sim}, namely Cosine similarity~\cite{singhal2001modern}, Euclidean distance~\cite{duda1973pattern} and Manhattan~\cite{krause1986taxicab} distance. \crp{To convert all measures to a consistent [0,1] range, we transform cosine similarity via \(\frac{\mathrm{cosine\_similarity} + 1}{2}\), and map distance measures from \([0,+\infty)\) to \([0,1]\) by defining \(\mathrm{sim}(\mathbf{x}, \mathbf{y}) = \exp\bigl( -\lambda\,d(\mathbf{x},\mathbf{y})\bigr)\).} As $d(\mathbf{x}, \mathbf{y})$ increases, the exponential term approaches 0, thus reducing similarity towards 0. Conversely, for smaller distances, $\mathrm{sim}(\mathbf{x}, \mathbf{y})$ approaches 1. We learn the IPSsim \(\alpha\in[0,1]\) (in increments of 0.1) on a validation set, and test for \(|T|\in\{5,10\}\). In each top-n scenario, we mitigate large inverse propensity values using a clipping function 
\(\tau_{top-n} = \tfrac{10}{\sqrt{Number of clicks}}\) (e.g., \(\tau_{top-n}=0.1\) when number of clicks \(=10^4\)), 
promoting stable learning outcomes.

\noindent \textbf{Click Model:} Following Oosterhuis~\cite{oosterhuis2023doubly}, we (1) use the affine click model to simulate user clicks, (2) when calculating the position bias, we set the probability of a document being observed by a user, $P(O = 1 \mid k)$, as $\left(1 + \frac{(k - 1)}{5}\right)^{-2}$, we set the probability that a user will click a document if they think that the document is relevant as 1, and the probability that the user will click the document if they think that the document is not relevant as $0.1 + \frac{0.6}{(1 + k/20)}$, and (3) we normalize the relevance scores, $rel(i)$, in the range $[0,1]$.  

\noindent \textbf{Logging policy:} As our historical ranking model (often referred to as a \textit{logging policy}), we follow Oosterhuis~\cite{oosterhuis2023doubly} and deploy a neural network with two hidden layers, each with 32 units that are optimised via policy gradients estimated using the PL-Rank-2~\cite{oosterhuis2021computationally}, trained on 1\% of the annotated training data. When simulating clicks, we uniformly sample queries associated with the historical ranking generated from the logging policy. We deploy three top-n settings, top-10, top-30 and top-50, and three different numbers of training clicks, $10^4$, $10^6$ and $10^8$.

\looseness -1\noindent \textbf{Metrics and Significance Test:} In Section~\ref{sec:results}, each approach is evaluated over five independent runs, and we report the mean NDCG~\cite{jarvelin2002cumulated} scores for each approach. As our test of statistical significance, we use a Student's t-test, with $p < 0.05$, over the five runs. In Tables~\ref{tab:tab1} and~\ref{tab:tab2}, an underlined value denotes the best performing IPSsim estimator across all six proposed IPSsim estimators, while a bold value indicates the best performance among all of the evaluated estimators. An upward black triangle ($\blacktriangle$) denotes a significant difference and better performance than the underlined estimator, while the downward black triangle ($\blacktriangledown$) denotes a significant difference and worse performance than the underlined estimator.

\begin{table*}[ht]
  \caption{NDCG of top-n settings on ISTELLA-S. $\blacktriangle$ and $\blacktriangledown$ indicates statistically significant increase and decrease ($p < 0.05$), \underline{underline} indicates the optimal performance of IPSsim with a specific $\alpha$ value (0.1 to 0.9).}
  \renewcommand{\arraystretch}{1} 
  \Large
  \resizebox{1\linewidth}{!}{%
  \begin{tabular}{l|lHlHl|lHlHl|lHlHl}
    \toprule
    & \multicolumn{5}{c}{Number of clicks = $10^4$} & \multicolumn{5}{c}{Number of clicks = $10^6$} & \multicolumn{5}{c}{Number of clicks = $10^8$}\\
    \midrule
    top-n & n = 10 & n = 20 & n = 30 & n = 40 & n = 50 & n = 10 & n = 20 & n = 30 & n = 40 & n = 50 & n = 10 & n = 20 & n = 30 & n = 40 & n = 50\\
    \midrule
    LP & 0.6797 & 0.7482 & 0.7807 & 0.7939 & 0.8007 & 0.6797 & 0.7482 & 0.7807 & 0.7939 & 0.8007 & 0.6797 & 0.7482 & 0.7807 & 0.7939 & 0.8007\\
    \midrule
    Naive & 0.6797 & 0.7482$^{\blacktriangledown}$ & 0.7808$^{\blacktriangledown}$ & 0.7938$^{\blacktriangledown}$ & 0.8007$^{\blacktriangledown}$ & 0.6797$^{\blacktriangledown}$ & 0.7482$^{\blacktriangledown}$ & 0.7806$^{\blacktriangledown}$ & 0.7939$^{\blacktriangledown}$ & 0.8006$^{\blacktriangledown}$ &  0.6826$^{\blacktriangledown}$ & 0.7482$^{\blacktriangledown}$ & 0.7846$^{\blacktriangledown}$ & 0.7939$^{\blacktriangledown}$ & 0.8047$^{\blacktriangledown}$\\
    IPS &0.6797 & 0.7482$^{\blacktriangledown}$ & 0.7806$^{\blacktriangledown}$ & 0.7939$^{\blacktriangledown}$ & 0.8008$^{\blacktriangledown}$ & 0.7220 & 0.7976 & 0.8086$^{\blacktriangledown}$ & 0.8148$^{\blacktriangledown}$ & 0.8244$^{\blacktriangledown}$ & 0.7343 & 0.8006 & 0.8203$^{\blacktriangledown}$ & 0.8230 & 0.8316 \\
    RPS & \textbf{0.7030}$^{\blacktriangle}$ & 0.6657$^{\blacktriangledown}$   & 0.7212$^{\blacktriangledown}$   & 0.6547$^{\blacktriangledown}$   & 0.6635$^{\blacktriangledown}$ & 0.7182  & 0.5957$^{\blacktriangledown}$   & 0.4499$^{\blacktriangledown}$   & 0.3788$^{\blacktriangledown}$   & 0.7825$^{\blacktriangledown}$ & 0.7213$^{\blacktriangledown}$   & 0.5145$^{\blacktriangledown}$   & 0.4178$^{\blacktriangledown}$  & 0.3905 & 0.7841$^{\blacktriangledown}$ \\
    DR & 0.7023$^{\blacktriangle}$ & \textbf{0.7761} & 0.8034$^{\blacktriangledown}$ & 0.8056$^{\blacktriangledown}$ & 0.8096$^{\blacktriangledown}$& \textbf{0.7267} & 0.7844 & 0.8048$^{\blacktriangledown}$  & 0.8128$^{\blacktriangledown}$ & 0.8135$^{\blacktriangledown}$ & \textbf{0.7429} & 0.7972 & 0.8170$^{\blacktriangledown}$ & 0.8274$^{\blacktriangledown}$ & 0.8277$^{\blacktriangledown}$ \\
    MIPS\textsubscript{LTR} &0.6798  & 0.7482$^{\blacktriangledown}$  & 0.7807$^{\blacktriangledown}$  & 0.7939$^{\blacktriangledown}$  & 0.8154 & 0.7078$^{\blacktriangledown}$  & 0.7880 & 0.8127$^{\blacktriangledown}$ & 0.8181 & 0.8236$^{\blacktriangledown}$ & 0.7106$^{\blacktriangledown}$  & 0.7887$^{\blacktriangledown}$  & 0.8191$^{\blacktriangledown}$  & 0.8216$^{\blacktriangledown}$  & 0.8301\\
    DRM & 0.6915$^{\blacktriangle}$ & 0.7610  & 0.8007$^{\blacktriangledown}$  & 0.7939  & 0.8127 & 0.3173$^{\blacktriangledown}$  & 0.6671  & 0.7988$^{\blacktriangledown}$  & 0.8229  & 0.8254 & 0.3511$^{\blacktriangledown}$  & 0.4074  & 0.6094$^{\blacktriangledown}$  & 0.8037 & 0.8325 \\
    \midrule
    IPSsim${(cosa_5)}$ & 0.6797$_{(0.6)}$ & \underline{0.7483}$_{(0.1)}$ & 0.8026$_{(0.9)}$ & 0.8069$_{(0.1)}$ & \underline{\textbf{0.8224}}$_{(0.1)}$ & 0.7164$_{(0.9)}$ & 0.7985$_{(0.9)}$ & 0.8248$_{(0.7)}$ & \underline{\textbf{0.8340}}$_{(0.8)}$ & \underline{\textbf{0.8344}}$_{(0.9)}$ & \underline{0.7396}$_{(0.1)}$ & 0.8034$_{(0.2)}$ & 0.8259$_{(0.9)}$ & \underline{\textbf{0.8342}}$_{(0.8)}$ & 0.8352$_{(0.8)}$\\
    IPSsim${(euca_5)}$ & 0.6797$_{(0.3)}$ & 0.7483$_{(0.1)}$ & 0.7807$_{(0.1)}$ & 0.8014$_{(0.6)}$ & 0.8092$_{(0.7)}$ & \underline{0.7246}$_{(0.5)}$ & 0.7898$_{(0.9)}$ & 0.8164$_{(0.9)}$ & 0.8279$_{(0.9)}$ & 0.8306$_{(0.9)}$ & 0.7371$_{(0.2)}$ & 0.8007$_{(0.4)}$ & 0.8219$_{(0.9)}$ & 0.8292$_{(0.9)}$ & 0.8316$_{(0.3)}$\\
    IPSsim${(man_5)}$ & \underline{0.6798}$_{(0.5)}$ & 0.7482$_{(0.4)}$ & 0.7807$_{(0.3)}$ & 0.8011$_{(0.5)}$ & 0.8206$_{(0.6)}$ & 0.7226$_{(0.6)}$ & \underline{\textbf{0.8061}}$_{(0.2)}$ & \underline{\textbf{0.8280}}$_{(0.3)}$ & 0.8264$_{(0.9)}$ & 0.8321$_{(0.9)}$ & 0.7282$_{(0.4)}$ & \underline{\textbf{0.8062}}$_{(0.4)}$ & 0.8197$_{(0.9)}$ & 0.8245$_{(0.6)}$ & 0.8325$_{(0.9)}$\\
    IPSSim${(cosa_{10})}$ & 0.6797$_{(0.6)}$ & 0.7482$_{(0.3)}$ & 0.7807$_{(0.5)}$ & \underline{\textbf{0.8181}}$_{(0.8)}$ & 0.8007$_{(0.9)}$ & 0.7126$_{(0.9)}$ & 0.7928$_{(0.9)}$ & 0.8224$_{(0.9)}$ & 0.8217$_{(0.9)}$ & 0.8319$_{(0.7)}$ & 0.7210$_{(0.9)}$ & 0.7886$_{(0.9)}$ & \underline{\textbf{0.8277}}$_{(0.8)}$ & 0.8316$_{(0.2)}$ & \underline{\textbf{0.8399}}$_{(0.7)}$\\
    IPSsim${(euca_{10})}$ & 0.6797$_{(0.2)}$ & 0.7482$_{(0.3)}$ & \underline{\textbf{0.8148}}$_{(0.2)}$ & 0.7939$_{(0.2)}$ & 0.8007$_{(0.9)}$ & 0.7195$_{(0.7)}$ & 0.7984$_{(0.9)}$ & 0.8218$_{(0.9)}$ & 0.8333$_{(0.9)}$ & 0.8312$_{(0.9)}$ & 0.7189$_{(0.9)}$ & 0.7911$_{(0.9)}$ & 0.8227$_{(0.8)}$ & 0.8340$_{(0.9)}$ & 0.8331$_{(0.9)}$\\
    IPSsim${(man_{10})}$ & 0.6797$_{(0.4)}$ & 0.7482$_{(0.1)}$ & 0.7807$_{(0.2)}$ & 0.7939$_{(0.3)}$ & 0.8007$_{(0.4)}$ & 0.7183$_{(0.7)}$ & 0.7900$_{(0.9)}$ & 0.8190$_{(0.9)}$ & 0.8271$_{(0.9)}$ & 0.8315$_{(0.9)}$ & 0.7193$_{(0.9)}$ & 0.7931$_{(0.9)}$ & 0.8243$_{(0.7)}$ & 0.8260$_{(0.9)}$ & 0.8320$_{(0.9)}$\\
    \midrule
  \end{tabular}}\label{tab:tab1}
\end{table*}

\section{Results}\label{sec:results}
\pageenlarge 1
Tables~\ref{tab:tab1} and~\ref{tab:tab2} present the performance of our three IPSsim
variants, IPSsim\textsubscript{(cosa)}, IPSsim\textsubscript{(euca)}, and IPSsim\textsubscript{(man)}, for \(T=5\) and \(T=10\). We compare IPSsim with the logging policy (LP), Naive, IPS, RPS, DR, MIPS\textsubscript{LTR}, and DRM approaches. Both tables show results under \(10^4\), \(10^6\), and \(10^8\) clicks for top-n settings where n\( \in \{10,30,50\}\). We also report the learned \(\alpha\) for each IPSsim variant.

From Table~\ref{tab:tab1} (ISTELLA-S), for n = 30 with \(10^4\), \(10^6\), and \(10^8\)
clicks, at least one IPSsim variant outperforms all other estimators; at n = 50, IPSsim
surpasses most baselines (including DR). From Table~\ref{tab:tab2} (MSLTR-WEB30K), for
n = 50 under \(10^6\) and \(10^8\) clicks, IPSsim again outperforms most baselines. Hence, we can answer \textbf{RQ1} positively: integrating document similarity into IPS improves
ranking effectiveness when n \(\ge 30\).

\begin{table*}[tb]
  \caption{NDCG of top-n settings on MSLTR-WEB30K dataset. $\blacktriangle$ and$\blacktriangledown$ indicates statistically significant increase and decrease ($p < 0.05$), \underline{underline} indicates the optimal performance of IPSsim with a specific $\alpha$ value (0.1 to 0.9).}
  \renewcommand{\arraystretch}{1} 
  \Large
  \resizebox{1\linewidth}{!}{%
  \begin{tabular}{l|lHlHl|lHlHl|lHlHl}
    \toprule
    & \multicolumn{5}{c}{Number of clicks = $10^4$} & \multicolumn{5}{c}{Number of clicks = $10^6$} & \multicolumn{5}{c}{Number of clicks = $10^8$}\\
    \midrule
    top-n & n = 10 & n = 20 & n = 30 & n = 40 & n = 50 & n = 10 & n = 20 & n = 30 & n = 40 & n = 50 & n = 10 & n = 20 & n = 30 & n = 40 & n = 50 \\
    \midrule
    LP & 0.4734 & 0.4969 & 0.5224 & 0.5491 & 0.5756 & 0.4734 & 0.4969 & 0.5224 & 0.5491 & 0.5756 & 0.4734 & 0.4969 & 0.5224 & 0.5491 & 0.5756 \\
    \midrule
    Naive & 0.4733 & 0.4969 & 0.5224 & 0.5490 & 0.5756 & 0.4827$^{\blacktriangledown}$ & 0.4968$^{\blacktriangledown}$ & 0.5224$^{\blacktriangledown}$ & 0.5490$^{\blacktriangledown}$ & 0.5756$^{\blacktriangledown}$ & 0.4781$^{\blacktriangledown}$ & 0.4998$^{\blacktriangledown}$ & 0.5224$^{\blacktriangledown}$ & 0.5491$^{\blacktriangledown}$ & 0.5756$^{\blacktriangledown}$ \\
    IPS & 0.4734 & 0.4969 & 0.5224 & 0.5491 & 0.5756 & 0.5099 & 0.5214 & 0.5388$^{\blacktriangledown}$ & 0.5558$^{\blacktriangledown}$ & 0.5756$^{\blacktriangledown}$ & 0.5221 & 0.5421 & \textbf{0.5596} & 0.5688 & 0.5793$^{\blacktriangledown}$  \\
    DR & \textbf{0.5054}$^{\blacktriangle}$ & 0.5349$^{\blacktriangle}$ & 0.5523$^{\blacktriangle}$ & \textbf{0.5791}$^{\blacktriangle}$ & 0.5756 & \textbf{0.5224}$^{\blacktriangle}$ & \textbf{0.5396}$^{\blacktriangle}$ & \textbf{0.5586}$^{\blacktriangle}$ & 0.5665 & 0.5773$^{\blacktriangledown}$ & \textbf{0.5397}$^{\blacktriangle}$ & \textbf{0.5475} & 0.5518 & 0.5688 & 0.5804$^{\blacktriangledown}$  \\
    MIPS\textsubscript{LTR} &0.4734 & 0.4968 & 0.5225 & 0.5491 & 0.5756& 0.4887 & 0.5145 & 0.5403 & 0.5491$^{\blacktriangledown}$ & 0.5756$^{\blacktriangledown}$& 0.4868$^{\blacktriangledown}$ & 0.5131$^{\blacktriangledown}$ & 0.5224$^{\blacktriangledown}$ & 0.5491$^{\blacktriangledown}$ & 0.5756$^{\blacktriangledown}$  \\
    DRM & 0.5000$^{\blacktriangle}$ & 0.5255$^{\blacktriangle}$  & \textbf{0.5538}$^{\blacktriangle}$  & 0.5491  & 0.5756  & 0.4033$^{\blacktriangledown}$  & 0.3968$^{\blacktriangledown}$  & 0.4023$^{\blacktriangledown}$  & 0.4374$^{\blacktriangledown}$  & 0.5057$^{\blacktriangledown}$  & 0.3290$^{\blacktriangledown}$  & 0.3543$^{\blacktriangledown}$  & 0.3608$^{\blacktriangledown}$  & 0.3968$^{\blacktriangledown}$ & 0.4069$^{\blacktriangledown}$  \\
    \midrule
    IPSsim${(cosa_{5})}$ & \underline{0.4734}$_{(0.7)}$ & \underline{0.4970}$_{(0.9)}$ & \underline{0.5225}$_{(0.2)}$ & \underline{0.5491}$_{(0.3)}$ & 0.5755$_{(0.9)}$ & 0.4734$_{(0.4)}$ & 0.4969$_{(0.4)}$ & 0.5224$_{(0.7)}$ & \underline{\textbf{0.5679}}$_{(0.1)}$ & \underline{\textbf{0.5898}}$_{(0.3)}$ & 0.5001$_{(0.8)}$ & 0.5248$_{(0.8)}$ & 0.5566$_{(0.6)}$ & \underline{\textbf{0.5807}}$_{(0.2)}$ & 0.5756$_{(0.4)}$ \\
    IPSsim${(euca_{5})}$ & 0.4734$_{(0.2)}$ & 0.4969$_{(0.1)}$ & 0.5224$_{(0.4)}$ & 0.5491$_{(0.7)}$ & \underline{0.5756}$_{(0.3)}$ & 0.4956$_{(0.9)}$ & \underline{0.5239}$_{(0.9)}$ & \underline{0.5473}$_{(0.9)}$ & 0.5491$_{(0.5)}$ & 0.5756$_{(0.3)}$ & 0.5078$_{(0.9)}$ & 0.5309$_{(0.9)}$ & 0.5510$_{(0.9)}$ & 0.5490$_{(0.1)}$ & 0.5755$_{(0.2)}$ \\
    IPSsim${(man_{5})}$ & 0.4734$_{(0.3)}$ & 0.4970$_{(0.5)}$ & 0.5225$_{(0.8)}$ & 0.5490$_{(0.3)}$ & 0.5756$_{(0.4)}$ & \underline{0.4995}$_{(0.9)}$ & 0.4969$_{(0.1)}$ & 0.5224$_{(0.6)}$ & 0.5490$_{(0.2)}$ & 0.5756$_{(0.1)}$ & \underline{0.5124}$_{(0.9)}$ & \underline{0.5323}$_{(0.9)}$ & 0.5462$_{(0.7)}$ & 0.5490$_{(0.1)}$ & \underline{\textbf{0.5919}}$_{(0.1)}$ \\
    IPSSim${(cosa_{10})}$ & 0.4734$_{(0.4)}$ & 0.4970$_{(0.1)}$ & 0.5224$_{(0.4)}$ & 0.5490$_{(0.8)}$ & 0.5756$_{(0.3)}$ & 0.4944$_{(0.9)}$ & 0.4968$_{(0.4)}$ & 0.5224$_{(0.7)}$ & 0.5490$_{(0.7)}$ & 0.5756$_{(0.6)}$ & 0.4940$_{(0.9)}$ & 0.4969$_{(0.9)}$ & 0.5224$_{(0.9)}$ & 0.5490$_{(0.8)}$ & 0.5756$_{(0.8)}$ \\
    IPSsim${(euca_{10})}$ & 0.4734$_{(0.2)}$ & 0.4969$_{(0.1)}$ & 0.5224$_{(0.7)}$ & 0.5490$_{(0.4)}$ & 0.5756$_{(0.8)}$ & 0.4904$_{(0.9)}$ & 0.5152$_{(0.9)}$ & 0.5224$_{(0.3)}$ & 0.5491$_{(0.4)}$ & 0.5756$_{(0.2)}$ & 0.4940$_{(0.9)}$ & 0.5233$_{(0.8)}$ & 0.5225$_{(0.8)}$ & 0.5491$_{(0.3)}$ & 0.5756$_{(0.3)}$ \\
    IPSsim${(man_{10})}$ & 0.4734$_{(0.5)}$ & 0.4969$_{(0.7)}$ & 0.5224$_{(0.9)}$ & 0.5491$_{(0.7)}$ & 0.5756$_{(0.6)}$ & 0.4992$_{(0.9)}$ & 0.5250$_{(0.9)}$ & 0.5225$_{(0.1)}$ & 0.5490$_{(0.6)}$ & 0.5756$_{(0.2)}$ & 0.5121$_{(0.9)}$ & 0.5446$_{(0.6)}$ & \underline{0.5572}$_{(0.9)}$ & 0.5817$_{(0.8)}$ & 0.5756$_{(0.2)}$ \\
    \midrule
  \end{tabular}}\label{tab:tab2}
\end{table*}

Additionally, the observation that IPSsim is particularly effective in top-n settings when n $\geq$ 30 provides insights into \textbf{RQ2}, and we can conclude that top-n settings do indeed affect the performance of our IPSsim estimator (compared to the other evaluated estimators). The DR estimator initially performs better than our IPSsim estimators in smaller top-n settings on both datasets ISTELLA-S and MSLTR-WEB30K. However, as the size of n increases, we find that our IPSsim estimators often perform better than most existing IPS estimators. We also find that although different similarity calculation methods do not affect the performance of our IPSsim estimator, on both of the datasets, IPSsim$_{(cosa)}$ often outperforms other IPSsim estimators. 

In response to \textbf{RQ3}, we observe that \zl{as the number of training clicks increases from \( 10^4 \) to \( 10^6 \) to \( 10^8 \), our IPSsim estimators show improved performance in mitigating position bias as measured by NDCG.} Therefore, we conclude that as the number of training clicks increases, we see improved performance when deploying IPSsim.

\begin{figure}[ht]
    \centering
    \includegraphics[width=0.8\linewidth, height=0.4\linewidth]{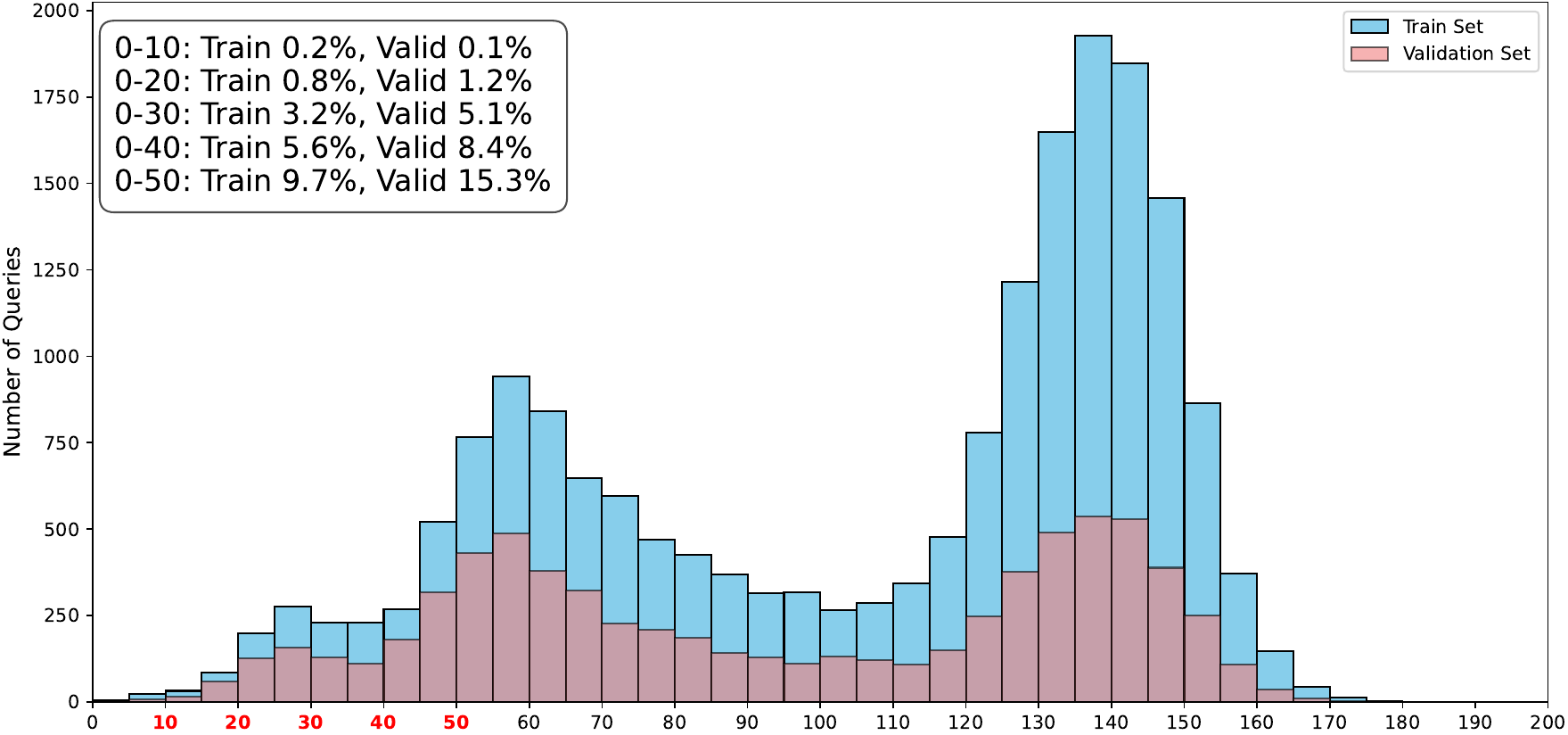}
    \caption{Distribution of Queries by Document Count on Dataset ISTELLA-S (Train and Validation).}
    \label{fig:ISTELLA-label}
\end{figure}

\begin{figure}[ht]
    \centering
    \includegraphics[width=0.8\linewidth, height=0.4\linewidth]{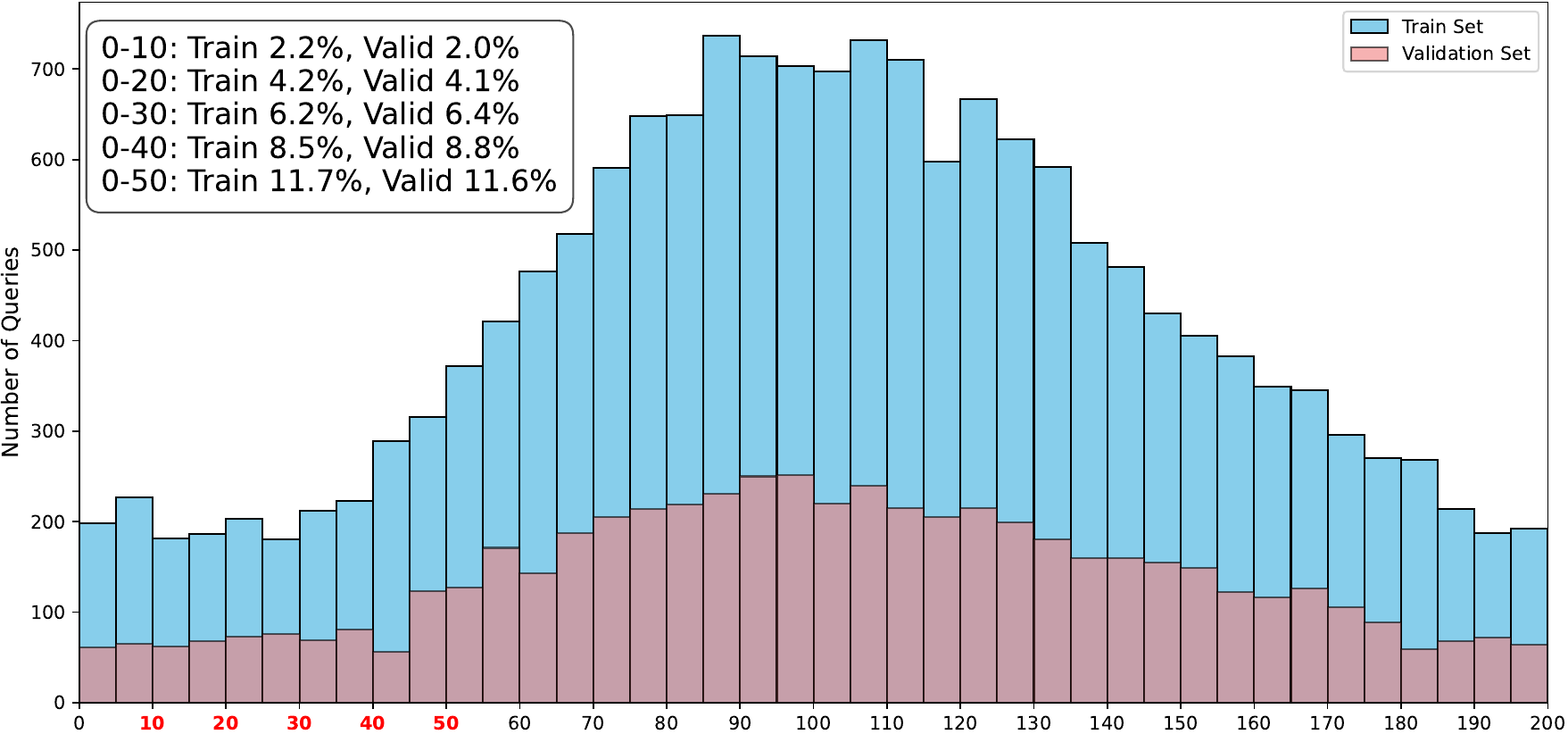}
    \caption{Distribution of Queries by Document Count on Dataset MSLTR-WEB30K (Train and Validation).}
    \label{fig:msltr-label}
\end{figure}
\pageenlarge 1
On ISTELLA-S, IPSsim boosts the propensity scores of lower-ranked but similar documents, effectively addressing position bias by integrating document similarity into the IPS framework. When the number of clicks is \(10^6\) or \(10^8\), $IPSsim$ surpasses most of the baselines, highlighting the benefits of leveraging semantic similarity in debiasing. Among the baselines, each method shows moderate performance overall. However, in certain settings, all baselines except the naive approach can slightly outperform the other baseline estimators. Overall, as we increase n on the MSLTR-WEB30K dataset, the improvements in debiasing are less pronounced for all estimators compared to on the ISTELLA-S dataset. To better understand this pattern, we conducted a distribution analysis to explore possible causes and guide potential improvements.
\pageenlarge 1
We show the number of documents per query in \crp{Figures~\ref{fig:ISTELLA-label}} and\crp{~\ref{fig:msltr-label}}. \crp{W}e also examine the distribution of annotated relevance labels in the ISTELLA-S and MSLTR-WEB30K datasets to understand why ULTR methods perform differently.
In the MSLTR-WEB30K dataset, many queries contain too few candidate documents, limiting the click data \crp{that} ULTR methods rely on.
Additionally, MSLTR-WEB30K has fewer documents with the highest relevance labels but a large number of non-relevant ones, leading to label imbalance and increased noise. By contrast, ISTELLA-S queries tend to have more candidate documents and better coverage of high-relevance labels, allowing ULTR methods to achieve more robust bias mitigation and higher performance overall.

\section{Conclusion}
In this study, we proposed IPSsim for extending the IPS calculation in ULTR, such that the calculation takes into consideration the similarity of low-ranked documents to highly ranked relevant documents. Our experiments on two standard LTR datasets, with different top-n settings, showed that our IPSsim estimator improves LTR performance, particularly when n >= $30$. For example, when \zl{n = 30} and \zl{number of clicks = \( 10^6 \)}, our IPSsim estimator achieves a 2.88\% improvement ($p < 0.05$) in terms of \crp{NDCG} compared to the Doubly Robust estimator from the literature. As future work, we will adapt our IPSsim ULTR approaches to modern transformer-based IR models, as well as explore the effectiveness of IPSsim ULTR approaches on datasets that contain the click interactions of real users.

\balance


\end{document}